\documentclass[a4paper,11pt]{article}
\usepackage{pos}
\usepackage{subcaption}
\usepackage{ifthen}
\usepackage{tikz}

\newcommand{\GeV}{{\rm GeV}}
\newcommand{\TeV}{{\rm TeV}}

\newcommand{\lrb}[1]{\left( #1 \right)}

\newcommand{\lrBigb}[1]{\Big( #1 \Big)}
\newcommand{\lrBigsb}[1]{\Big[ #1 \Big]}

\newcommand{\lrBiggsb}[1]{\Bigg[ #1 \Bigg]}

\newcounter{NumArgs}

\newcommand{\eqs}[1]{\setcounter{NumArgs}{0}\foreach\i in{#1}{\stepcounter{NumArgs}}%
\ifthenelse{\equal{\theNumArgs}{1}}{eq.~(\ref{#1})}%
{\ifthenelse{\equal{\theNumArgs}{2}}%
{eqs.~\foreach\i[count=\q]in{#1}{\ifthenelse{\equal{\q}{\theNumArgs}}{and (\ref{\i})}{(\ref{\i})~}}}%
{eqs.~\foreach\i[count=\q]in{#1}{\ifthenelse{\equal{\q}{\theNumArgs}}{and (\ref{\i})}{(\ref{\i}),~}}}}}

\newcommand{\Eqs}[1]{\setcounter{NumArgs}{0}\foreach\i in{#1}{\stepcounter{NumArgs}}%
\ifthenelse{\equal{\theNumArgs}{1}}{Eq.~(\ref{#1})}%
{\ifthenelse{\equal{\theNumArgs}{2}}%
{Eqs.~\foreach\i[count=\q]in{#1}{\ifthenelse{\equal{\q}{\theNumArgs}}{and (\ref{\i})}{(\ref{\i})~}}}%
{Eqs.~\foreach\i[count=\q]in{#1}{\ifthenelse{\equal{\q}{\theNumArgs}}{and (\ref{\i})}{(\ref{\i}),~}}}}}

\newcommand{\refs}[1]{\setcounter{NumArgs}{0}\foreach\i in{#1}{\stepcounter{NumArgs}}%
\ifthenelse{\equal{\theNumArgs}{1}}{(\ref{#1})}%
{\ifthenelse{\equal{\theNumArgs}{2}}%
{\foreach\i[count=\q]in{#1}{\ifthenelse{\equal{\q}{\theNumArgs}}{and (\ref{\i})}{(\ref{\i})~}}}%
{\foreach\i[count=\q]in{#1}{\ifthenelse{\equal{\q}{\theNumArgs}}{and (\ref{\i})}{(\ref{\i}),~}}}}}

\newcommand{\Figs}[1]{\setcounter{NumArgs}{0}\foreach\i in{#1}{\stepcounter{NumArgs}}%
\ifthenelse{\equal{\theNumArgs}{1}}{Fig.~(\ref{#1})}%
{\ifthenelse{\equal{\theNumArgs}{2}}%
{Figs.~\foreach\i[count=\q]in{#1}{\ifthenelse{\equal{\q}{\theNumArgs}}{and (\ref{\i})}{(\ref{\i})~}}}%
{Figs.~\foreach\i[count=\q]in{#1}{\ifthenelse{\equal{\q}{\theNumArgs}}{and (\ref{\i})}{(\ref{\i}),~}}}}}

\newcommand{\Gen}[2]{\setcounter{NumArgs}{0}\foreach\i in{#2}{\stepcounter{NumArgs}}%
	\ifthenelse{\equal{\theNumArgs}{1}}{#1.~(\ref{#2})}%
	{\ifthenelse{\equal{\theNumArgs}{2}}%
		{#1.~\foreach\i[count=\q]in{#2}{\ifthenelse{\equal{\q}{\theNumArgs}}{and (\ref{\i})}{(\ref{\i})~}}}%
		{#1.~\foreach\i[count=\q]in{#2}{\ifthenelse{\equal{\q}{\theNumArgs}}{and (\ref{\i})}{(\ref{\i}),~}}}}}

\title{Leptogenesis and the relativistic degrees of freedom of the plasma}
\author*[a,b]{Dimitrios Karamitros}
\author[c]{Thomas McKelvey}
\author[c,d]{Apostolos Pilaftsis}

\affiliation[a]{Helsinki Institute of Physics, University of Helsinki,\\
 P.O. Box 64, FIN-00014, Helsinki, Finland}

\affiliation[b]{Department of Physics, University of Jyv\"askyl\"a,\\
  P.O.Box 35 (YFL), FIN-40014,  Jyv\"askyl\"a, Finland}

\affiliation[c]{Department
	of Physics and Astronomy, University of Manchester,\\
	M13 9PL,  Manchester, United Kingdom}
	
\affiliation[d]{PRISMA Cluster of Excellence \& Mainz Institute for Theoretical Physics,\\ Johannes Gutenberg University, 55099 Mainz, Germany}

\emailAdd{dimitrios.d.karamitros@jyu.fi}
\emailAdd{thomas.mckelvey@manchester.ac.uk}
\emailAdd{apostolos.pilaftsis@manchester.ac.uk}

\abstract{
	We investigate the impact of the temperature dependence of the {\em relativistic degrees of freedom} (dofs) of the plasma on lepton and baryon asymmetry. 
	Motivated by the significant effect of the varying dofs on the {\em tri-resonant leptogenesis} particle model in low-scale leptogenesis, we show how this effect impacts the evolution of the lepton asymmetry in a simplified setup.
	We provide analytical approximations as well as numerical results showing that the simplified setup exhibits similar behavior as the concrete model.
	As the dofs enter the transport equations via the expansion rate of the Universe and the temperature of the plasma, we argue that any  analysis must take these effects into account in order to be consistent.
}

\FullConference{Proceedings of the Corfu Summer Institute 2024 "School and Workshops on Elementary Particle Physics and Gravity" (CORFU2024)\ 12 - 26 May, and 25 August - 27 September, 2024\\ Corfu, Greece\\}

\begin{document}
\maketitle

\section{Introduction}
\label{sec:intro}
The matter-antimatter asymmetry of the Universe provides evidence of physics beyond the Standard Model (SM), with observations (e.g. by Planck~\cite{Planck:2018vyg}) measuring the Baryon Asymmetry of the Universe (BAU) to be $\eta_B \approx 6 \times 10^{-10}$.

It is well established that the three Sakharov conditions~\cite{Sakharov:1967dj} are a necessary ingredient for the production of BAU. In addition, leptogenesis~\cite{FukYan:1986} offers a minimal explanation to BAU, by only extending the SM by heavy particles that seed lepton asymmetry, which is converted into a baryon asymmetry through $(B+L)$-violating sphaleron transitions~\cite{KUZMIN198536}.

In many simple extensions, these additional particles have (CP-violating) Yukawa interactions with leptons, which can generate masses for the neutrinos via the {\em seesaw} mechanism~\cite{Minkowski:1977sc,Yanagida:1979as,Mohapatra:1979ia,GellMann:1980vs,Schechter:1980gr}, providing an additional motivation for particle models of baryogenesis through leptogenesis.

In this contribution, we focus on how the temperature dependence of the effective relativistic degrees of freedom (dofs) affects the evolution of the lepton asymmetry, since such effects are typically assumed to be negligible. 
As a concrete example, we summarize the findings of our previous works on a class of leptogenesis models~\cite{daSilva:2022mrx,Karamitros:2023tqr} in which we first observed these effects. 
To understand their origin and provide a more intuitive picture, we focus on a simplified version of the transport equations in which we can obtain analytical approximations for the evolution of the neutrino and lepton asymmetry densities. 
This is a simple exercise that shows how the temperature dependence of the dofs affects the baryon asymmetry.  

The text is organized as follows. 
In Section~\ref{sec:TRL}, we show the results of the so-called ``tri-resonant leptogenesis'' (TRL) class of models~\cite{daSilva:2022mrx,Karamitros:2023tqr}.  
In Section~\ref{sec:simple}, we study a simplified set of evolution equations for the heavy-neutrino and lepton asymmetry densities, providing both results from their numerical integration and  analytical approximations.
In Section~\ref{sec:sum}, concisely present our conclusions.

\section{A concrete realization: Tri-Resonant Leptogenesis}
% show the difference due to h_eff
%
\label{sec:TRL}
\subsection{The setup of ``Resonant Leptogenesis''}
Due to the lightness of the SM neutrinos, it is typical to expect heavy-neutrino masses at the scale of Grand Unified Theory (GUT), which is a typical issue in leptogenesis. As a result, since the interactions between the light and heavy neutrino species are suppressed by the heavy-neutrino masses, the experimental detection becomes improbable.
An elegant solution to this issue is provided by the framework of \textit{Resonant Leptogenesis} (RL)~\cite{Pilaftsis:1997jf,Pilaftsis:2003gt,Pilaftsis:2005rv}, in which the CP asymmetry is enhanced through the mixing of near-degenerate heavy-neutrinos that satisfy
\begin{equation}
	\left| m_{N_\alpha} - m_{N_\beta} \right|\, \simeq\, \frac{1}{2}\Gamma_{\alpha,\beta}\;.
	\label{eq:resonance_condition}
\end{equation}
Here, $m_{N_\alpha}$ and $\Gamma_\alpha$ are the mass and decay width of the heavy neutrino species $N_\alpha$, respectively. 
It is noteworthy that RL can achieve the observed BAU with $ m_{N_\alpha}$ at sub-$\TeV$ scales whilst maintaining agreement with the measurements of neutrino oscillation parameters, providing a natural and testable framework for both BAU and neutrino mass problems. 

\subsection{The Tri-Resonant Leptogenesis class of models}
\begin{figure}
	\centering
	\includegraphics[width=0.4\linewidth]{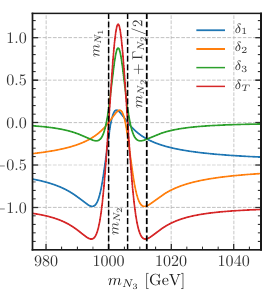}
	\caption{The parameter $\delta_{\alpha}$ for each $N_{\alpha}$, as a function of the heaviest neutrino mass, $m_{N_3}$. The values of $m_{N_{1,2}}$ are chosen to satisfy the resonance condition~(\ref{eq:resonance_condition}). The parameter $\delta_T$ is the sum of all, $\delta_T = \sum\limits_{\alpha}\delta_{\alpha}$.}
	\label{fig:deltaCP}
\end{figure}
In the TRL, introduced in ref.~\cite{daSilva:2022mrx} and further studied in ref.~\cite{Karamitros:2023tqr}, the CP asymmetry is maximized through constructive interference of all heavy neutrinos, which allows for the baryon asymmetry to be generated with larger-scale Yukawa couplings. 

The relevant TRL Lagrangian terms are 
\begin{equation}\label{eq:Lagr}
	-\mathcal{L}^{\nu_R} = \boldsymbol{h}^{\nu}_{ij} \overline{L}_i \tilde{\Phi} \nu_{R,j} + \frac{1}{2}\overline{\nu}_{R,i}^C \lrb{\boldsymbol{m}_M}_{ij} \nu_{R,j} + {\rm H.c.} \; ,
\end{equation}
where $L_i = \lrb{\nu_{i\, L},\, e_{i\, L}}^{\sf T}$ the left-handed ${\rm SU(2)}_L$ lepton doublets, and $\tilde{\Phi} = i\sigma_2\Phi^*$ the weak-isospin-conjugate Higgs doublet.

It can be shown~\cite{daSilva:2022mrx}  that the SM neutrinos acquire masses given by
\begin{equation}
	\boldsymbol{m}^\nu = - \frac{v^2}{2 m_N} \lrb{\boldsymbol{h}^{\nu}(\boldsymbol{h}^\nu)^{\sf T} + \mathcal{O}\lrb{\frac{\Delta \boldsymbol{m}_M}{m_N} }}\, .
\end{equation}
Hence, for singlet neutrinos with a near-degenerate mass spectrum, the SM neutrino mass matrix can approximately vanish by demanding 
\begin{equation}
	\boldsymbol{h}^{\nu}(\boldsymbol{h}^\nu)^{\sf T} = \boldsymbol{0}_3 \, .
\end{equation}
This motivates the central assumption of TRL
\begin{equation}\label{eq:Yukawa}
	\boldsymbol{h}^\nu_0 = \begin{pmatrix}
		a & a\, \omega & a\,\omega^2\\
		b & b\,\omega & b\,\omega^2\\
		c & c\,\omega & c\,\omega^2
	\end{pmatrix}\; ,
\end{equation}
where $a,\, b, \, c \, \in \mathbb{C}$, and $\omega$ are the generators of the discrete group $\mathbb{Z}_6$ (or $\mathbb{Z}_{3}$). This simple symmetric Yukawa structure suppresses the SM neutrino masses without requiring GUT-scale $m_{N}$, while providing a CP asymmetry that is enough to explain BAU observations.

The production of BAU is mostly proportional to $\delta_{1,2,3}$ (see e.g.~\cite{Deppisch:2010fr,daSilva:2022mrx}), which quantify the CP-asymmetry generated by the CP violating decays of the heavy-neutrinos, $N_{1,2,3}$.  For concreteness, in \Figs{fig:deltaCP} we show the typical values of $\delta_{1,2,3}$ that we find viable SM neutrino masses and BAU. It is noteworthy that  maximum of $\delta_T = \sum\limits_{\alpha}\delta_{\alpha}$ close or above unity is generic in TRL, at any mass scale, $m_{N_1}$, as long as all masses satisfy the resonance condition~(\ref{eq:resonance_condition}).

\subsection{Leptogenesis and the varying dofs in TRL}
In our  works on the TRL, we followed several previous analyses (e.g.~\cite{Sigl:1993ctk,BhupalDev:2014pfm,Pilaftsis:2005rv}) to derive the semi-classical Boltzmann equations~\cite{daSilva:2022mrx} as well as the flavor covariant transport equations~\cite{Karamitros:2023tqr} that take into account the contributions from coherent neutrino oscillation.~\footnote{The physics behind the evolution equations is pivotal in understanding the details of the particle and QFT view of leptogenesis, but it is out of the scope of this proceedings contribution.} 
The main difference between our works and previous analyses, is the consistent inclusion of all terms associated with the temperature dependence of the dofs, and the careful analysis of their contribution to the evolution of both the neutrino and lepton asymmetry densities.

\begin{figure}[t]
	\hspace{0.5cm}
	\includegraphics[width=0.7\linewidth]{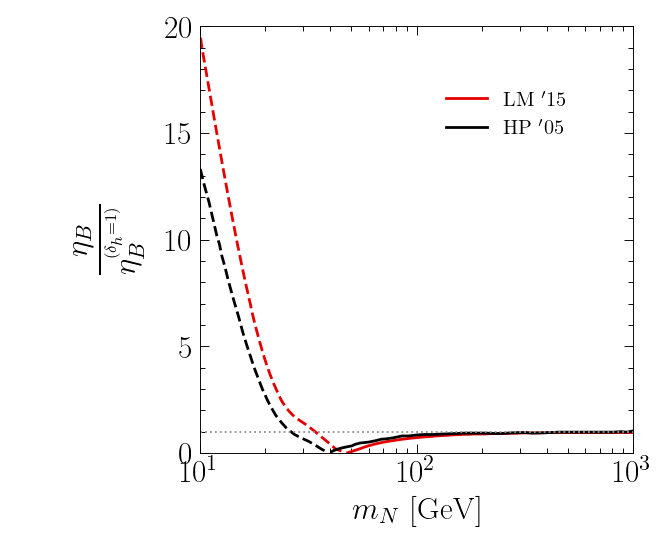}
	\caption{The ratio of the BAU produced in TRL between varying ($\eta_B$) and constant $\lrb{\eta_B^{(\delta_h=1)}}$ dofs as a function of the neutrino mass scale,  $m_N$. The two lines correspond to different computations of the temperature depended quantities of the plasma; with red and black corresponding to Laine and Meyer~\cite{Laine:2015kra} and  Hindmarsh and Philipsen~\cite{Hindmarsh:2005ix}, respectively. Also, solid (dashed) lines show a positive (negative) ratio.}
	\label{fig:etaB_ratio_TRL}
\end{figure}
Our results in both analyses can be summarized in \Figs{fig:etaB_ratio_TRL}, which shows the ratio of $\eta_B$ with varying ($\eta_B$) and constant $\lrb{\eta_B^{(\delta_h=1)}}$ dofs as a function of the neutrino mass scale,  $m_N$, extracted from ref.~\cite{Karamitros:2023tqr}. The red and black lines correspond two different computations~\cite{Laine:2015kra,Hindmarsh:2005ix} of plasma thermodynamic quantities (e.g. pressure, energy, and entropy densities), while the solid (dashed) lines indicate a positive (negative) ratio.
As can be seen, for $m_N \lesssim 100~\GeV$ the difference becomes significant, with $m_N \approx 40~\GeV$ being a turning point in which $\eta_B$ goes to zero. For $m_N \lesssim 40~\GeV$ the ratio also exhibits a sign change, something that does not occur if we do not consider varying dofs.
It is worth mentioning that neglecting the coherent neutrino oscillations (as in ref.~\cite{daSilva:2022mrx}) yields  almost identical ratio, including the turning point at $m_N \approx 40~\GeV$ and the sign changed for $m_N \lesssim 40~\GeV$.

\begin{figure}[t]
%	\centering
	\begin{subfigure}{0.52\linewidth}
		%\centering
		\hspace{-0.3cm}
		\includegraphics[width=\linewidth]{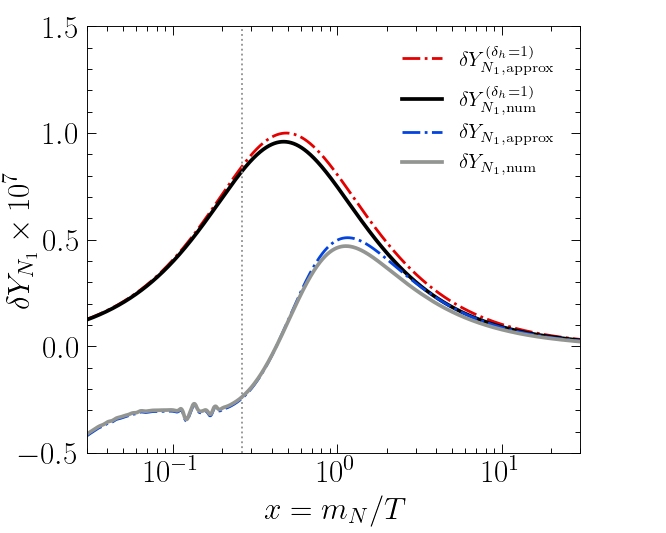}
		\caption{ }
		\label{fig:deltaYN_approx_vs_num}
	\end{subfigure}
	\begin{subfigure}{0.52\linewidth}
		%\centering
%		\hspace{-1.5cm}
		\includegraphics[width=\linewidth]{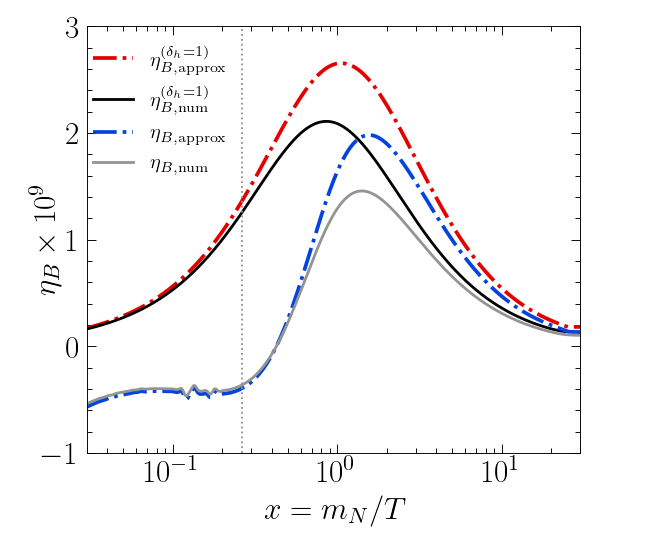}
		\caption{\empty}
		\label{fig:etaB_approx_vs_num}
	\end{subfigure}
	\caption{The right (left) panel shows the evolution of the baryon asymmetry (neutrino deviation from equilibrium) in TRL for $m_N =35~\GeV$ and $|\boldsymbol{h}^{\nu}| \sim 10^{-4}$. The solid lines correspond to numerrical solution of the transport equations, while the dot-dashed lines correspond to the attractive semi-analytical approximations. In gray and blue (black and red) are show these quantities for varying (constant) dofs. The vertical line shows a typical choice for the sphaleron decoupling temperature, $T_{\rm sph} = 132~\GeV$.}
	\label{fig:approx_vs_num}
\end{figure}
We point out that the results we observe do not seem to be attributed to numerical errors or instabilities, as we have employed a number of different methods for solving the transport equations that include explicit, semi-implicit, and implicit Runge-Kutta methods provided by several available tools~\cite{NaBBODES,2020SciPy-NMeth}. 
Moreover, due to the well-documented attractive nature of the transport equations (see, e.g.~\cite{Pilaftsis:2005rv,Deppisch:2010fr,daSilva:2022mrx}), we are able to approximately solve for the evolution of both the $N_1$ deviation from equilibrium ($\delta Y_{N}= \delta\eta_N$) and lepton-asymmetry without relying on numerical methods, just by requiring that the right-hand-side of the transport equations nearly vanishes. This is exemplified in \Figs{fig:approx_vs_num}, which shows that the fully numerical solutions and the approximations are in agreement.~\footnote{We only show $\delta Y_{N_1}$, but the behavior is similar for the other neutrinos as well.}

\section{Impact of varying dgreess of freedom is generic}
%show that the simplified equations are affected the same way
%
\label{sec:simple}
In this section we aim to shed some light on the results we obtained for TRL, by modeling the complete set of transport equations with a simple system that consists of a neutrino and a lepton component. To make the evolution as simple as possible, we only include the neutrino decay rate ($\Gamma_N$) -- including the real-intermediate part of the $2 \to 2$ scatterings needed~\cite{Kolb:1979qa} to obey the Sakharov equilibrium condition~\cite{Sakharov:1967dj} -- and the associated CP-violation ($\delta_N$). Thus, for the sake of simplicity, we neglect terms that include lepton back-reactions, neutrino coherent oscillations, and scatterings. Such terms can affect the quantitative results, but this simple system has the elements needed to capture the quantitative picture of leptogenesis. 
\subsection{Simple set of equations}
The equations that govern the evolution of this system is 
\begin{eqnarray}
	&&\dfrac{d Y_N}{d x} =
	 - \delta_h \ \dfrac{K_1(x)}{K_2(x)} \dfrac{\Gamma_N}{x H(x)}  \lrBigb{ Y_N - Y_N^{\rm eq} } 
	 \label{eq:dYdx}
	 \\
	 &&\dfrac{d \eta_L}{d x} =  \dfrac{\Gamma_N}{H(x)}   \dfrac{K_1(x)}{2 \, x \, \zeta(3)} 
	 \lrBigb{ \delta_{N} \ \delta Y_{N} -2/3  \ \eta_L} - 3\dfrac{\eta_L}{x}(\delta_h -1) \;,
	 \label{eq:detaLdx}
\end{eqnarray} 
where $x=m_N/T$, $Y_N = n_N/s$ ($s$ is the entropy density) and $Y_N^{\rm eq}$ its equilibrium value, $\delta Y_N = \delta\eta_N =  Y_N/Y_N^{\rm eq} -1$, and $\delta_h = 1 + \dfrac{1}{3} \dfrac{d \ln h_{\rm eff}}{\ln T}$ ($h_{\rm eff} $ is the dofs as defined through $s$). Notice that the equations present in the literature  typically set $\delta_h =1$, corresponding to constant $h_{\rm eff}$. 
These equations can be solved numerically. However, we note that $Y_N$ -- and its difference from $Y_N^{\rm eq}$ -- can acquire extremely small values that could cause numerical instabilities and ``round off'' errors. Moreover, we note that what drives the lepton-asymmetry production is $\delta Y_N$. Therefore, instead of solving directly \eqs{eq:dYdx}, we may express it in a way that is  better suited for numerical computations as well as physically meaningful:
\begin{eqnarray}
	\dfrac{d \delta Y_N}{d x} =
	-  \dfrac{K_1(x)}{K_2(x)}  \lrBigsb{ 1+ \lrb{ 1 - \dfrac{\Gamma_N}{x\,H(x)} }\delta Y_N }
	-3  \dfrac{\delta Y_N+1}{x} (\delta_h - 1)
	\label{eq:ddeltaYdx}
\end{eqnarray}

\subsection{Analytical approximations}
\subsubsection{Approximation close to the initial condition}
\begin{figure}[t]
	\centering
	\includegraphics[width=0.6\linewidth]{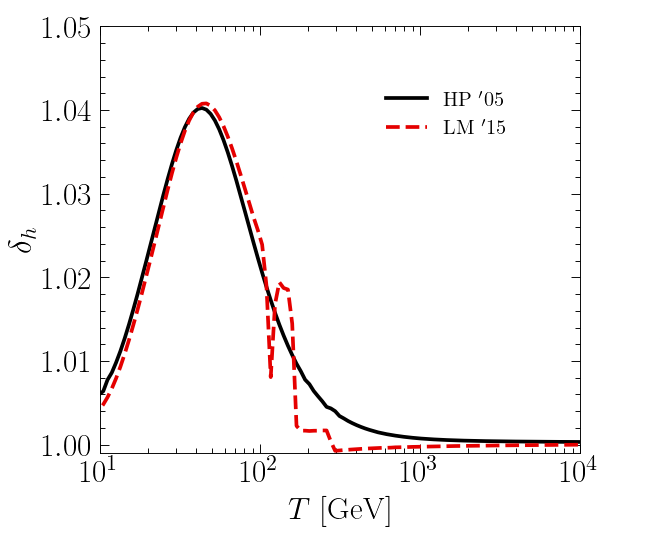}
		\caption{the values of $\delta_h$ of two tabulations provided by   Hindmarsh and Philipsen~\cite{Hindmarsh:2005ix} (black solid line) and Laine and Meyer~\cite{Laine:2015kra} (red dashed line). }
	\label{fig:deltah}
\end{figure}

Before solving numerically \eqs{eq:ddeltaYdx,eq:detaLdx}, it would be helpful to show the analytical approximations we obtain under some circumstances. Focusing on initial conditions $\eta_L=\delta Y_{N}=0$  at $x \to 0$, we can express  \eqs{eq:ddeltaYdx,eq:detaLdx} as
\begin{eqnarray}
	&&\dfrac{d \delta Y_N}{d x} \approx \dfrac{x}{2} - 3 (\delta_h - 1)
	\label{eq:ddeltaYdx_init}
	\\
	&&\dfrac{d \eta_L}{d x} \approx  \dfrac{\Gamma_N}{H(x)}   \dfrac{\delta_{N} }{2 \, x^2 \, \zeta(3)} \delta Y_{N}  \;.
	\label{eq:detaLdx_init}
\end{eqnarray} 
These equations show that, initially, $\delta Y_N$ can obtain both positive and negative values depending on the magnitude of $\delta_h$. In \Figs{fig:deltah}, we show the temperature dependence of $\delta_h -1$ computed by two different groups~\cite{Hindmarsh:2005ix,Laine:2015kra}. These two computations differ numerically, but they both show an increase at $T \approx 100~\GeV$. which can push $\delta Y_N$ to obtain negative values. Furthermore, since $\dfrac{d \eta_L}{d x} \sim \delta Y_{N}$, the sign of  $\delta Y_{N}$ directly affects whether initially $\eta_L$ will be positive or negative. 
\begin{figure}[t]
	\centering
	\includegraphics[width=0.6\linewidth]{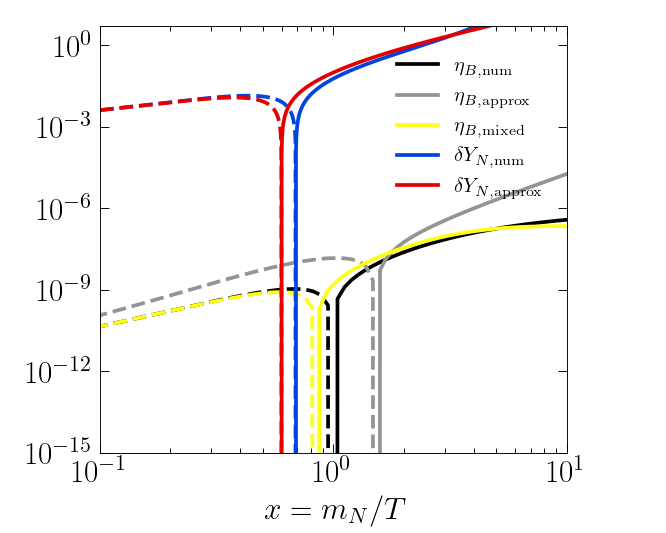}
	\caption{The evolution of $\eta_B$ and $\delta Y_{N}$ for $m_N = 35~\GeV$, $\Gamma_N=10^{-21}~\GeV$, and $\delta_N = -1$. The numerical solution for $\eta_B$ ($\delta Y_{N}$) is shown in black (blue), while the approximation in gray (red). The yellow line corresponds to using the analytical approximation of $\delta Y_{N}$ and numerically integrating eq.~(\ref{eq:detaLdx}). Solid (dashed) lines correspond to positive (negative) values.}
	\label{fig:init_approx}
\end{figure}
This behavior is reflected in the approximate solutions of \eqs{eq:ddeltaYdx_init,eq:detaLdx_init},
\begin{eqnarray}
	&& \delta Y_N \approx \dfrac{x^2-x_0^2}{2} + \ln\lrb{ \dfrac{h_{\rm eff}(x)}{h_{\rm eff}(x_0)}  }
	\label{eq:deltaY_init}
	\\
	&&\eta_L \approx  \dfrac{\Gamma_N}{H|_{T=m_N}}   \dfrac{\delta_{N}}{2 \zeta(3)} 
	\lrb{x-x_0}\lrBiggsb{ \dfrac{\lrb{x-x_0}\lrb{x+2x_0}}{12}  +\ln\lrb{ \dfrac{h_{\rm eff}(x)}{h_{\rm eff}(x_0)}  } }
	  \;.
	\label{eq:etaL_init}
\end{eqnarray} 
We note that we keep $x_0 \ll 1$ as the point at which we set the initial condition. Also, to obtain an analytical solution for $\eta_L$, we have to assume constant dofs. However, since the main impact comes from $\delta Y_{N}$, the approximation~\eqs{eq:etaL_init} captures the relevant behavior.

In \Figs{fig:init_approx} we show the evolution of $\delta Y_{N}$ and $\eta_B = - \dfrac{28}{51} \eta_L$. In this figure, approximation~\refs{eq:deltaY_init} is in good agreement with the numerical solution of \eqs{eq:ddeltaYdx}. Due to the assumption of constant dofs in obtaining \eqs{eq:etaL_init}, the numerical values between the approximation and the numerical solution of \eqs{eq:detaLdx} deviate, but \eqs{eq:etaL_init} still shows similar qualitative behavior as the numerical integration of \eqs{eq:detaLdx}. To properly take into account the varying dofs, we can use the approximation~\refs{eq:deltaY_init} to numerically integrate \eqs{eq:detaLdx} (yellow line), which results in better agreement with the fully numerical result.
We note that the  approximations~\refs{eq:deltaY_init, eq:etaL_init} rely on relatively low values of $\Gamma_N$. Increasing it, pushes $\delta Y_N $ and $\eta_L$ to their attractive solutions mentioned previously. 

\subsubsection{Attractive solution}
\begin{figure}[t]
	\centering
	\includegraphics[width=0.6\linewidth]{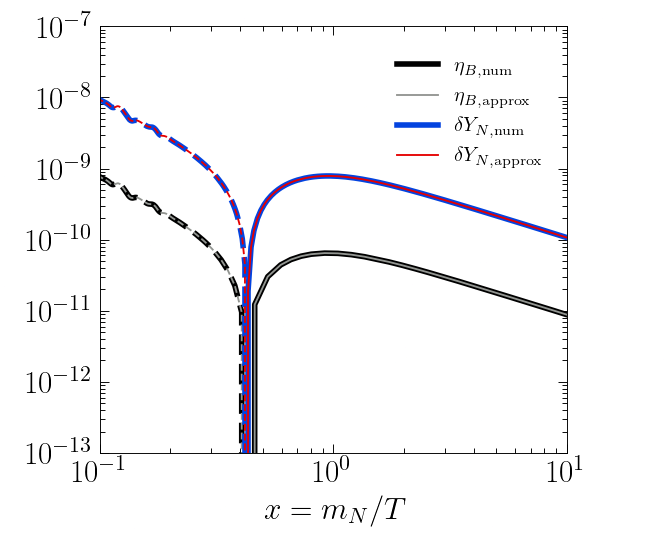}
	\caption{The evolution of $\eta_B$ and $\delta Y_{N}$ for $m_N = 35~\GeV$, $\Gamma_N=10^{-6}~\GeV$, and $\delta_N = -0.1$. The numerical solution for $\eta_L$ ($\delta Y_{N}$) is shown in black (blue), while the approximation in gray (red). Solid (dashed) lines correspond to positive (negative) values.}
	\label{fig:fixed_approx}
\end{figure}
The attractive solutions of \eqs{eq:ddeltaYdx,eq:detaLdx} are obtained by demanding that their right-hand-sides remain close to zero. The attractive solutions are reached if the system has enough time to allow the derivative to change sign. Once this happens, any change in $\eta_L$ or $\delta Y_N$ causes the corresponding derivative to change sign in the opposite direction. This behavior is extensively discussed in the literature (e.g.~\cite{Pilaftsis:2005rv,Deppisch:2010fr,daSilva:2022mrx}), and results in an independence from the initial conditions in most situations.  
The attractive solutions of \eqs{eq:ddeltaYdx,eq:detaLdx} are 
\begin{eqnarray}
	&& \delta Y_N \approx  
	\dfrac{ \dfrac{K_1(x)}{K_2(x)}  x+ 3(1- \delta_h)} {\lrb{\dfrac{\Gamma_N}{x H}-1}\dfrac{K_1(x)}{K_2(x)} x  -3  (1- \delta_h)}
	\label{eq:deltaY_fixed}
	\\[0.5cm]
	&&\eta_L \approx \dfrac{3}{2}\dfrac{\dfrac{\Gamma_N K_1(x)}{3\zeta(3)H} }{\dfrac{\Gamma_N K_1(x)}{3\zeta(3)H}   +3   (1-\delta_h)} \ \delta_N   \ \delta Y_N
	\;.
	\label{eq:etaL_fixed}
\end{eqnarray} 
These approximations again show the possibility to have both positive and negative $\delta Y_{N}$, which directly affects the sign of $\eta_L$. \Figs{fig:fixed_approx} demonstrates a case in which both  $\delta Y_{N}$ and $\eta_B =  - \dfrac{28}{51} \eta_L$ change sign during their evolution, and also the  remarkable agreement between the attractive solutions and  the numerical integration.

\subsection{Numerical results}
\begin{figure}
	\hspace{0.5cm}
	\includegraphics[width=0.7\linewidth]{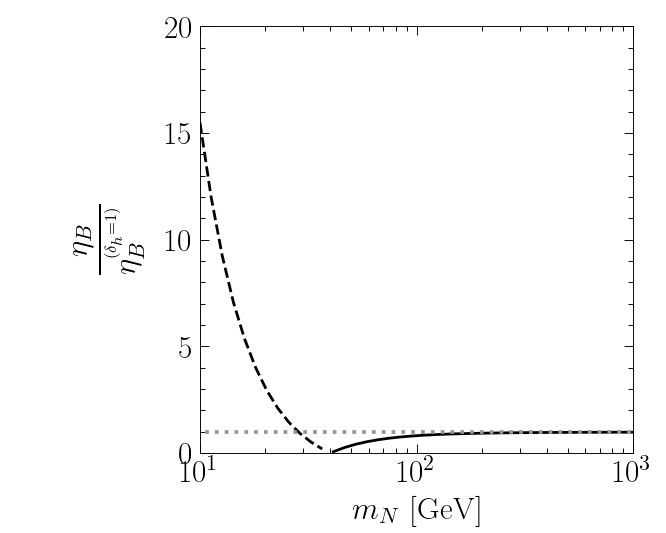}
	\caption{The ratio of the BAU produced (in this simplified system) between varying ($\eta_B$) and constant $\lrb{\eta_B^{(\delta_h=1)}}$ dofs as a function of the neutrino mass scale,  $m_N$, for  $\Gamma_N=4\times 10^{-8} \  m_N$ and $\delta_{N}=1$. The result is obtained using the dofs computed in ref.~\cite{Hindmarsh:2005ix}. The solid (dashed) line denote positive (negative) ratio.}
	\label{fig:etaB_ratio_simple}
\end{figure}
The approximations presented offer a better understanding on the origin of the impact the varying dofs, especially $\delta_h$, on the evolution of both $\delta Y_{N}$ and $\eta_L$. Moreover, they both show that $\eta_L$ agrees with Sakharov conditions, as it is proportional to $\delta_{N} \times \delta Y_{N}$.

In \Figs{fig:etaB_ratio_simple}, we show  the BAU generated using different values of $m_N$. In complete analogy to the TRL case, \Figs{fig:etaB_ratio_TRL}, we observe that the impact of the varying dofs ($\delta_h \neq 1$) becomes significant for $m_N \lesssim 100~\GeV$, with a sign change at $m_N \approx 40~\GeV$.
This implies that the impact of the temperature dependence of the dofs is model-independent. This is not surprising, since the dofs enter the transport equations regardless of the interactions between heavy-neutrinos and leptons. 
%
%In fact, similar effects are important in WIMP dark matter~\cite{Hindmarsh:2005ix}, but at the QCD scale $T~\sim 1 \GeV$. Since the QCD scale is mostly irrelevant for leptogenesis, the dofs around electroweak scale turn out to be of importance.

\begin{figure}[t]
	%	\centering
	\begin{subfigure}{0.52\linewidth}
		%\centering
		\hspace{-0.3cm}
		\includegraphics[width=\linewidth]{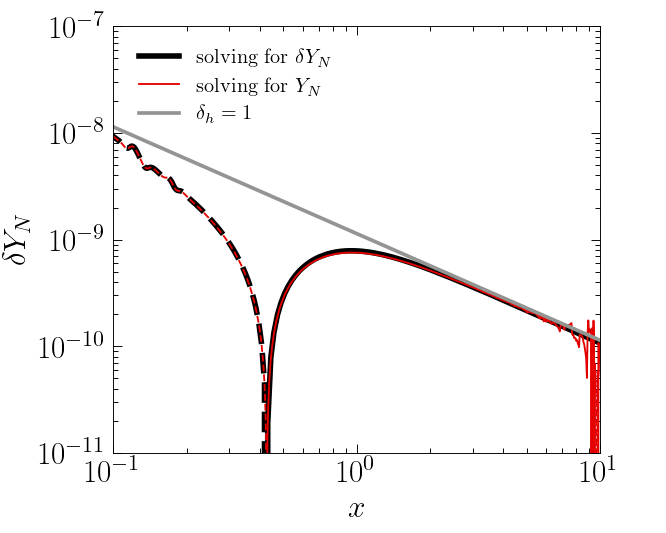}
		\caption{ }
		\label{fig:Y_compare}
	\end{subfigure}
	\begin{subfigure}{0.52\linewidth}
		%\centering
		%		\hspace{-1.5cm}
		\includegraphics[width=\linewidth]{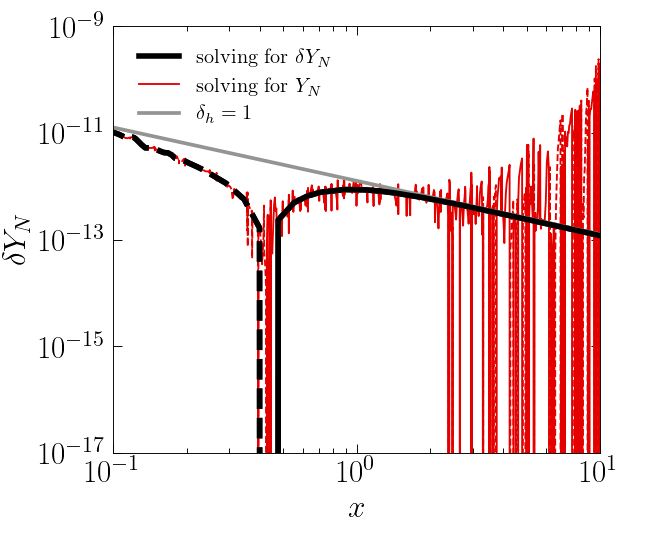}
		\caption{\empty}
		\label{fig:Y_instabilities}
	\end{subfigure}
	\caption{The evolution of  $\delta Y_{N}$ obtained by solving eq.~(\ref{eq:ddeltaYdx}) (black) and eq.~(\ref{eq:dYdx}) (red), and setting constant dofs (gray) for $m_N=35~\GeV$ with  \textbf{a)} $\Gamma_N=10^{-6}~\GeV$ and \textbf{a)} $\Gamma_N=10^{-3}~\GeV$}
	\label{fig:solve_Y_vs_deltaY}
\end{figure}

As a final remark, we would like to emphasize that \eqs{eq:dYdx,eq:ddeltaYdx} are completely equivalent. However, they may behave differently when integrated numerically, because the former can exhibit undesirable numerical artifacts.
In \Figs{fig:solve_Y_vs_deltaY}, we compare the evolution of  $\delta Y_{N}$ obtained by solving \eqs{eq:ddeltaYdx} (black) and \eqs{eq:dYdx} (red).~\footnote{To produce these figures, we use the ``BDF'' method~\cite{doi:10.1073/pnas.38.3.235} found in the {\tt solve\_ivp} module of {\tt scipy} with {\tt atol} = {\tt rtol} = $10^{-13}$. These are extreme values for {\tt atol} and {\tt rtol}, but lower values have a hard time stabilizing \eqs{eq:dYdx}. Similarly, the BDF method is one of the few available methods that can solve \eqs{eq:dYdx}. One the other hand, \eqs{eq:ddeltaYdx} is better suited for more numerical methods and can achieve stability with lower values of  {\tt atol} and {\tt rtol}. } In \Figs{fig:Y_compare}, both solutions agree, with the latter exhibiting an unstable behavior for $x \gtrsim 10$. In \Figs{fig:Y_instabilities}, the two solutions agree only for $x \lesssim 0.3$, with the solution of \eqs{eq:dYdx} quickly becoming irregular. 
The only difference between \Figs{fig:Y_compare,fig:Y_instabilities} is the value of $\Gamma_N$, which indicates that the instability depends on the parameters of the model at hand. Therefore, apart from being a more physical choice, \eqs{eq:ddeltaYdx} is numerically friendly and produces more reliable results.

\section{Conclusions}
\label{sec:sum}
We have studied how the process of leptogenesis  is affected by the consistent inclusion of the temperature dependence of the relativistic degrees of freedom of the plasma. 

In Section~\ref{sec:TRL}, we have shown the impact in a concrete model that produces the observed BAU as well as naturally explain the SM neutrino masses. We presented that, although different computations of the dofs result in slightly different numerical results, the qualitative effect of the varying dofs persists. We also demonstrated that our numerical computations agree with semi-analytical estimates, indicating that the effect we observe is not caused by numerical instabilities.

In Section~\ref{sec:simple}, we focused on two evolution equations that model a simple neutrino-lepton system. We provided analytical approximations for the evolution of $\delta Y_N$ and $\eta_L$ that capture the behavior of numerical results while showing how the dofs affect them.
We showed that the ratio of the produced BAU between the varying and constant dofs follows the same pattern as our concrete TRL model, indicating that the impact of the varying dofs is model independent. 
We also showed that, although there are different ways we can express the evolution equations, we should be careful of the form we choose to use, as it may suffer from numerical instabilities.

In closing, we emphasize that this contribution builds on~\cite{daSilva:2022mrx,Karamitros:2023tqr} to provide a model-independent argument for the impact of the temperature dependence of the  dofs on the production of BAU. 
Arguably, we have shown that our previously obtained results seem to be intrinsic to the transport equations. Therefore, future analyses must include these effects regardless of the model, especially if the lepton asymmetry in produced close or below the electroweak scale.
As the precision of cosmological measurements is constantly improving -- such as the recent results from the ACT collaboration~\cite{ACT:2025fju} -- details regarding the transport equations become more relevant. Therefore, phenomenological analyses that consistently take into account all relevant effects, including the varying dofs, are bound to play an important role in our predictions and design of new experiments.

\section*{Acknowledgements} 
\noindent
The work of AP  is supported in part by the Lancaster-Manchester-Sheffield Consortium for Fundamental Physics, under STFC Research Grant ST/T001038/1. TM acknowledges support from the STFC Doctoral Training Partnership under STFC training grant ST/V506898/1.

%\begin{thebibliography}{99}
%\bibitem{...}
%....
%
%\end{thebibliography}
%
\bibliography{refs}{}
\bibliographystyle{JHEP}

\end{document}